\documentclass[aps,prd,preprintnumbers,twocolumn]{revtex4-1}

\usepackage{color}
\usepackage{graphicx}
\usepackage{dcolumn}
\usepackage{bm}
\usepackage{ulem}
\usepackage{appendix}
\usepackage{amsmath}
\usepackage[thicklines]{cancel}

\begin{document}
\preprint{RIKEN-iTHEMS-Report-25}

\title{HoloNet: Toward a Unified Einstein-Maxwell-Dilaton Framework of QCD}

\author{Hong-An Zeng}
\affiliation{$^a$Center for Theoretical Physics and College of Physics, Jilin University, Changchun 130012, China}
\email{zengha20@mails.jlu.edu.cn}

\author{Lingxiao Wang}
\affiliation{$^{b}$RIKEN Interdisciplinary Theoretical and Mathematical Sciences (iTHEMS), Wako, Saitama 351-0198, Japan}
\email{lingxiao.wang@riken.jp}
\affiliation{$^{c}$Institute for Physics of Intelligence, Graduate School of Science, The University of Tokyo, Bunkyo-ku, Tokyo 113-0033, Japan}

\author{Mei Huang}
\affiliation{$^{d}$School of Nuclear Science and Technology, University of Chinese Academy of Sciences, Beijing 100049, China}
\email{huangmei@ucas.ac.cn}


\begin{abstract}
We propose HoloNet, a neural-network framework that unifies lattice QCD(LQCD) thermodynamics and holographic Einstein-Maxwell-Dilaton (EMD) theory within a data-to-holography pipeline. Instead of assuming specific functional forms, HoloNet learns the metric profile $A(z)$ and the gauge–dilaton coupling $f(z)$ directly from 2+1-flavor LQCD data at $\mu=0$. These learned functions are embedded into the EMD equations, enabling the model to reproduce the lattice equation of state and baryon number fluctuations with high fidelity. Once trained, HoloNet provides a fully data-driven holographic description of QCD that extends naturally to finite density, allowing us to map the phase diagram and estimate the location of the critical end point (CEP). The reconstructed potential $V(\phi)$ and coupling $f(\phi)$ agree quantitatively with those obtained from holographic renormalization, demonstrating that HoloNet can consistently bridge different holographic models.

\end{abstract}

\onecolumngrid
\maketitle

\section{Introduction}
The Quantum Chromodynamics (QCD) phase structure plays a crucial role in understanding the evolution of the early universe as well as dense matter properties inside the compact stars. It is widely believed that, at vanishing chemical potential, a crossover transition occurs at a temperature $T_c$. When extended to finite chemical potential, this crossover evolves into a first-order phase transition line~\cite{schmidt2017phase}. The point where the crossover meets the first-order transition line is the critical point, which is of great experimental significance. The search for this critical point is a central goal of relativistic heavy-ion collision experiments at Relativistic Heavy Ion Collisions (RHIC)~\cite{aggarwal2010higher,aggarwal2010experimental,adamczyk2014energy,luo2017search,adam2021nonmonotonic,abdallah2023higher}, as well as one of the scientific goals of upcoming facilities such as Facility for Antiproton and Ion Research (FAIR), Nuclotron-based Ion Collider Facility (NICA), and High Intensity Heavy-ion Accelerator Facility (HIAF).

Due to the sign problem~\cite{Philipsen:2012nu,philipsen2013qcd,Aarts:2015tyj,Nagata:2021ugx}, lattice QCD(LQCD) calculations at finite chemical potential remain extremely challenging. To study the QCD phase structure in the strong-coupling regime, many effective approaches have been employed to investigate the QCD phase diagram, including the Functional Renormalization Group (FRG)~\cite{fu2020qcd, zhang2017functional}, Dyson–Schwinger equations (DSEs)~\cite{gao2016qcd,qin2011phase,shi2014locate,fischer2014phase}, the Random Matrix Model (RMM)~\cite{halasz1998phase}, the Nambu–Jona-Lasinio (NJL) model~\cite{nambu1961dynamical}, and its extension, the PNJL model~\cite{sun2023splitting, li2019kurtosis, sasaki2010qcd, mclerran2009quarkyonic}. 
The AdS/CFT correspondence~\cite{maldacena1999large,witten1998anti,gubser1998gauge,aharony2000large}, also known as the holographic principle, has been developed as a powerful framework to address problems involving strong coupling, including the hard-wall model~\cite{erlich2005qcd}, the soft-wall model~\cite{karch2006linear}, and the V-QCD model~\cite{gursoy2018holographic}, among others. Moreover, another bottom-up theoretical framework, the Einstein-Maxwell-Dilaton (EMD) model, has recently attracted considerable attentions. It incorporates a dilaton field that breaks conformal symmetry and a Maxwell gauge field that introduces a chemical potential, thereby extending the model to finite density. The model appears to be “just sufficient” for describing QCD-like systems and has been shown to capture the essential features of the QCD equation of state(EoS) successfully. Furthermore, it can be extended to include additional effects, such as external magnetic fields~\cite{cai2024neural}. 

There are generally three approaches to determining the model. The first approach, following the spirit of Gubser-type models, assumes specific functional forms for the potential $V(\phi)$ and the coupling function $f(\phi)$ directly at the level of the action. The parameters of these functions are then fixed by comparing the model-predicted EoS on the boundary with LQCD data~\cite{dewolfe2011holographic,gubser2008mimicking,grefa2021hot,he2024gravitational,cai2022probing,li2024holographic,fu2025revisiting}. The second approach focuses on fixing the bulk geometry $A(z)$. In this case, data are used to determine the metric warp factor $A(z)$ and the coupling function $f(z)$ along the holographic direction. The corresponding potential $V(\phi)$ and coupling function $f(\phi)$ are then reconstructed, also known as the potential reconstruction method~\cite{Yang:2014bqa,yang2015confinement,dudal2017thermal,fang2016chiral,li2011thermodynamics,chen2019criticality,chen2021gluodynamics,zhou2020thermodynamics,chen2024machine,li2021entanglement}. The third arises from the fact that the dilaton field couples directly to the bulk geometry. One may instead assume a dilaton profile $\phi(z)$ along the holographic direction, from which both the bulk geometry and the potential functions can be solved~\cite{li2013dynamical,chen2022dynamical}. 

At the same time, the rapid development of machine learning~\cite{aarts2025physics} has endowed holographic models with significantly enhanced representational capabilities. In particular, the neural network approach enables a more efficient and effective solution of ordinary differential equations, especially for models in which the equations of motion contain functional dependencies that are difficult to handle analytically or numerically using traditional methods~\cite{song2021ads,jeong2025ads,ahn2025holographic, chen2018neural,chen2025data,hashimoto2025machine,hashimoto2024unification,ahn2024deep,ahn2025deep}. By introducing neural networks along the holographic direction~\cite{Hashimoto:2020jug, akutagawa2020deep,hashimoto2021neural}, it becomes possible to use boundary data and reconstruct the bulk metric. Using this approach, we can construct holographic models driven directly by LQCD data.

Another key motivation for incorporating machine learning into holographic QCD models is to reduce the inherent arbitrariness in model construction. At present, different holographic models—or even the same model solved via different approaches—often yield different parameters, resulting in significant discrepancies in the reconstructed gravitational geometry. Beyond variations arising from the choice of LQCD data, much of this discrepancy stems from the a priori selection of functional forms within the models. By replacing these arbitrary choices with neural networks with minimal physics priors~\cite{aarts2025physics}, one can largely suppress the human-imposed priors, thereby achieving an un-biased holographic QCD model~\cite{hashimoto2018deep,hashimoto2022deriving}. This approach allows different solution methods to be incorporated within a unified framework, enabling more consistent and reliable predictions.

In this work, we present HoloNet, a neural network–based approach to holographic QCD potential reconstruction. Our method directly learns the metric components $A(z)$ and coupling functions $f(\phi(z))$ from LQCD data, without assuming any specific functional forms. With necessary physics constraints, e.g., AdS boundary conditions, Stefan–Boltzmann limit, etc., we train our neural networks using 2+1 flavor LQCD data for the EoS and baryon number susceptibility at zero baryon chemical potential, achieving a fully data-driven reconstruction of the potential functions.

This paper is organized as follows. We first provide a brief introduction to the potential reconstruction EMD model. We then describe how to construct the neural network algorithm within this framework and explain the incorporation of constraints. Finally, we present comparisons with LQCD data and offer a discussion of the critical point predictions.

\section{HOLOGRAPHIC EMD MODEL}
We consider a 5-dimensional Einstein-Maxwell-dilaton (EMD) action:
\begin{equation}
S_E=\frac{1}{16 \pi G_5} \int d^5 x \sqrt{-g}\left[R-\frac{f(\phi)}{4} F^2-\frac{1}{2} \partial_\mu \phi \partial^\mu \phi-V(\phi)\right].
\end{equation}
Here, $f$ denotes the coupling function, the gauge field $F^2$ term provides the boundary baryon number density, the $\phi$ field is responsible for breaking conformal symmetry, and $G_5$ denotes the five-dimensional Newton's constant.

We consider the following ansatz for the metric,
\begin{equation}
    d s^2=\frac{L^2 e^{2 A(z)}}{z^2}\left[-g(z) d t^2+\frac{d z^2}{g(z)}+d \vec{x}^2\right],
\end{equation}
where the metric component $A(z)$ is the warp factor. $z = 0$ corresponds to the asymptotic AdS boundary, and we set $L = 1$ in this work.
The equations of motion can be solved by imposing the following boundary conditions.
At the horizon $z=z_H$,
\begin{equation}
    A_t\left(z_H\right)=g\left(z_H\right)=0.
\end{equation}
$A_t$ represents the time component of the gauge field.
At the AdS boundary $z=0$,
\begin{equation}
    \begin{aligned}
A(0) & =-\sqrt{\frac{1}{6}} \phi(0), g(0)=1, \\
A_t(0) & =\mu+\rho' z^2+\cdots.
\end{aligned}
\end{equation}
Under the holographic duality, $\mu$ corresponds to the chemical potential in the boundary field theory, while $\rho'$ corresponds to a quantity related to the baryon number density.
The baryon number density can be calculated as:
\begin{equation}
\begin{aligned}
\rho & =\left|\lim _{z \rightarrow 0} \frac{\partial \mathcal{L}}{\partial\left(\partial_z A_t\right)}\right| \\
& =-\frac{1}{16 \pi G_5} \lim _{z \rightarrow 0}\left[\frac{\mathrm{e}^{A(z)}}{z} f(\phi) \frac{\mathrm{d}}{\mathrm{~d} z} A_t(z)\right] .
\end{aligned}
\end{equation}
$\mathcal{L}$ denotes the Lagrangian density.
When the metric functions $A(z)$ and coupling functions $f(\phi)$ are undetermined, the equations of motion can be formally expressed in the following integral form. Based on Ref.~\cite{Yang:2014bqa}, we have simplified the relevant expressions to make them more suitable for constructing the neural network,
\begin{equation}
    \begin{aligned}
\phi^{\prime}(z) & =\sqrt{-6\left(A^{\prime \prime}-A^{\prime 2}+\frac{2}{z} A^{\prime}\right)}, \\
A_t(z)&=\mu \frac{\int_{z_H}^z \frac{y}{e^A f} d y}{\int_{z_H}^0 \frac{y}{e^A f} d y} ,\\
g(z)&=1+\frac{1}{\int_0^{z_H} y^3 e^{-3 A} d y}\left[-\int_{0}^z y^3 e^{-3 A} d y\right.\\
&\left.+\left(\frac{\mu}{\int_0^{z_H} \frac{y}{e^{A}f} d y}\right)^2 \hat{G}\right], \\
V(z) & =-3 z^2 g e^{-2 A}\left[A^{\prime \prime}+3 A^{\prime 2}+\left(\frac{3 g^{\prime}}{2 g}-\frac{6}{z}\right) A^{\prime}\right.\\
&\left.-\frac{1}{z}\left(\frac{3 g^{\prime}}{2 g}-\frac{4}{z}\right)+\frac{g^{\prime \prime}}{6 g}\right],
\end{aligned}
\end{equation}
where the expression of $\hat{G}$ is
\begin{equation}
    \hat{G}=\left|\begin{array}{ll}
\int_0^{z_H} y^3 e^{-3 A} d y & \int_0^{z_H} y^3 e^{-3 A} d y \int_{0}^y \frac{x}{e^{A}f} d x \\
\int_{z_H}^z y^3 e^{-3 A} d y & \int_{z_H}^z y^3 e^{-3 A} d y \int_{0}^y \frac{x}{e^{A}f} d x
\end{array}\right|.
\end{equation}

We can easily obtain the temperature $T$, entropy density $s$, and baryon number density $\rho$.
\begin{equation}
    \begin{aligned}
    T=&\frac{1}{4 \pi} \frac{z_H^3 e^{-3 A\left(z_H\right)}}{\int_0^{z_H} y^3 e^{-3 A(z)} d y}\left[1-\left(\frac{\mu}{\int_0^{z_H} \frac{y}{e^{A }f} d y}\right)^2\right.\\
    &\quad \left(\int_0^{z_H} y^3 e^{-3 A} d y \cdot \int_0^{z_H} \frac{x}{e^A f} d x\right. \\
    &\quad \left.\left.-\int_0^{z_H} y^3 e^{-3 A} d y \int_0^y \frac{x}{e^A f} d x\right)\right],\\
    s=&\frac{e^{3 A\left(z_H\right)}}{4 G_5 z_H^3},\\
    \rho =& \frac{1}{16 \pi G_5} \frac{\mu}{\int_{0}^{z_H} \frac{y}{e^A f}dy}.\label{intform}
\end{aligned}
\end{equation}
Using thermodynamic relations, the free energy can be expressed as
\begin{equation}
    F=-\int s d T-\rho d \mu.
\end{equation}
The pressure is given by $p=-F$. It is worth noting that, in general holographic dualities, the free energy typically contains contributions from the AdS background and is therefore divergent\cite{henningson1998holographic,balasubramanian1999stress,de2001holographic}. To ensure thermodynamic consistency, it is usually necessary to regularize it so that the first law of thermodynamics holds. In this work, we implement a simplified regularization by appropriately choosing the lower limit of integration, which amounts to fixing the integration constant(the lower limit fixed at $T=100$ MeV). This is because we are more concerned with the relative variation of the free energy. And the entropy at low temperatures is numerically closer to zero. We expect this to have an effect equivalent to subtracting the divergent terms in the action. Meanwhile, the energy density can be obtained as
\begin{equation}
    \epsilon=-p+s T+\mu \rho .
\end{equation}
The second-order baryon number susceptibility is defined as
\begin{equation}
    \chi_2^B=\frac{1}{T^2} \frac{\partial \rho}{\partial \mu}.
\end{equation}

In the above model, the unknown independent parameters include the horizon $z_H$, the chemical potential $\mu$, metric component $A$, the coupling function $f$, and the five-dimensional Newton's constant $G_5$ \footnote{In which, $A$ and potential function $V$ are mutually determined and are not independent variables}. On the dual QCD side, the independent parameters are 
temperature $T$ and chemical potential $\mu$. Now, the question is: \textit{which parameter's variation induces the change of QCD temperature $T$?}

Both the variation of $z_H$ and the change of $ A(z) $ can cause changes in the temperature $T$. A reasonable choice is to let different horizon positions $z_H$  correspond to different temperatures $T$. This allows us to use the same function $A(z)$ at all temperatures. Optimizing a single function is always simpler than optimizing multiple ones. This is also why we did not reparameterize the horizon to $z_H = 1$ from the beginning.

So far, all independent variables have their own physical meanings: $z_H$ and $\mu$ provide the degrees of freedom for temperature and chemical potential, while the functions $A(z)$ and $f(z)$ determine the values at each point on the $(T,\mu)$ phase diagram. The parameter $G_5$ sets the thermodynamic limit, which will be discussed later.

It is also worth noting that we can determine the model from LQCD data at zero chemical potential and then extend it to finite chemical potential. This is because the baryon number susceptibility, defined as the second derivative of the thermodynamic potential with respect to the chemical potential, already encodes information about the finite-$\mu$ regime. In the model, this information is encoded in the coupling function $f(z)$, which can be determined through the baryon number susceptibility. Once $f(z)$ is fixed, it in turn determines the thermodynamic quantities at finite chemical potential.

\section{Neural-Network Approach}
\begin{figure}[!htbp]
\centering
\includegraphics[width=1.\linewidth]{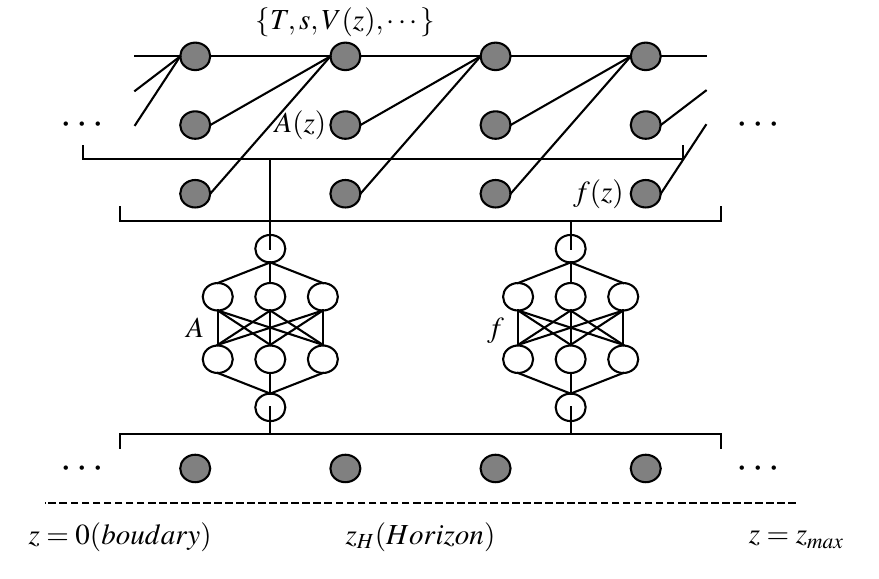}
\caption{\textbf{HoloNet}. A neural network is constructed along the holographic direction. Different bulk spacetimes are placed on a common holographic coordinate and differ only by their horizon locations $z_H$. The functions $A(z)$ and $f(z)$ are implemented as sub-networks (schematically shown as three-layer networks). Hollow nodes represent learnable parameters, while solid nodes represent the fixed equation of motion. The sub-networks output $A(z)$ and $f(z)$ at each layer and feed them into the fixed network to compute thermodynamic quantities.}
\label{fig:NN_structure}
\end{figure}
\textbf{HoloNet:} As mentioned above, our goal is to achieve a fully data-driven learning of the bulk geometry and its coupling to the gauge field. To this end, we employ deep neural networks to parameterize the functions $A$ and $f$. The network for $A(z)$ is a 4-layer fully connected architecture, \textit{1-12-tanh-23-tanh-12-tanh-1-(-softplus)}, and the network for $f(z)$ is \textit{1-12-tanh-23-tanh-1-softplus}, where the numbers denote the layer widths, \textit{tanh} is the hyperbolic activation, and \textit{softplus} is $\sigma(x)=\ln(1+e^x)$. Each network takes the holographic coordinate $z$ as input and outputs the corresponding value of $A(z)$ or $f(z)$. This architecture balances optimization efficiency with representational capacity. The outermost \textit{-softplus} in $A(z)$ facilitates imposing constraints involving derivative behaviors of $A$, while the final \textit{softplus} in $f(z)$ enforces its positivity. Further details will be provided later in the text.

\begin{figure*}[!htb]
    \centering
    \includegraphics[width=0.45\linewidth]{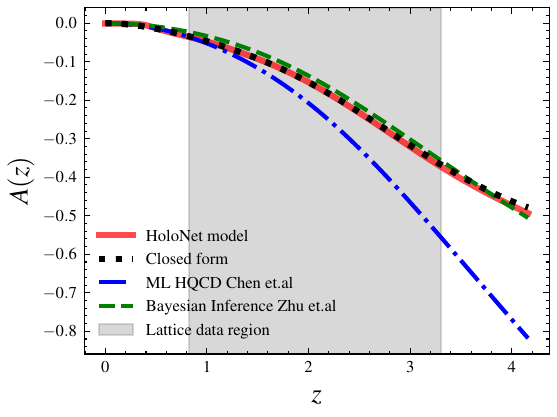}
    \includegraphics[width=0.45\linewidth]{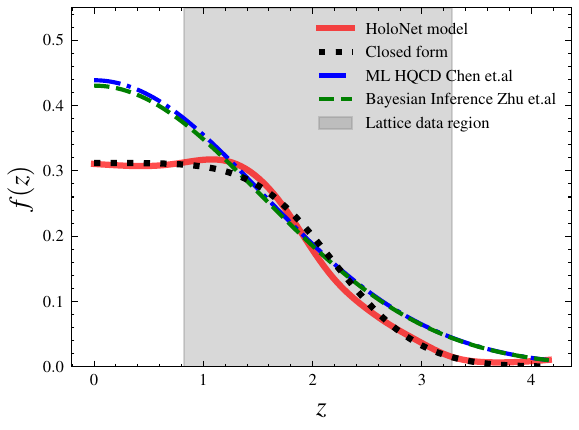}
    \caption{The graphs of the $A$ and $f$ functions are shown, with red representing the optimized HoloNet results, black corresponding to an analytically guessed solution, blue depicting the machine learning results obtained from a prescribed functional form~\cite{Chen:2024ckb}, and green indicating the outcomes of Bayesian analysis~\cite{zhu2025bayesian}. Additionally, the gray areas in the figures represent the $z$ range over which the lattice data are available.}
    \label{fig:Aandfform}
\end{figure*}

It is well known that neural networks can be used to solve differential equations~\cite{chen2018neural}. In our case, the equations of motion have already been reduced to the integral form~(\ref{intform}), and, as discussed above, the bulk geometries of different black hole solutions can be reparameterized into a unified metric (with the temperature encoded in $z_H$). This allows the general differential-equation network to be simplified into the structure shown in Fig.~\ref{fig:NN_structure}.

Our construction has three key components: 1) \textbf{Nesting networks}. The model consists of a large global network—corresponding to the integral form of the equations of motion—with fixed structure and coefficients, together with two independent sub-networks representing the functions $A(z)$ and $f(z)$; 2) \textbf{Automatic differentiation}. Because different black hole solutions share the same $A(z)$, they can be aligned on a common holographic coordinate. In the overlapping region, both the fields and metric take identical numerical values, and different spacetimes differ only by their respective $z_H$. This stacking enables efficient evaluation of thermodynamic quantities and their gradients using automatic differentiation; 3) \textbf{Self-adaptive optimization}. The sub-networks dynamically generate the layer-wise values of $A(z)$ and $f(z)$. Previous approaches~\cite{hashimoto2018deep} treat these values as direct parameters of the global network, reconstructing the geometry through their optimization. By contrast, our model computes them on the fly through explicit neural networks. Although seemingly more elaborate, this design is essential. The key reason is that the relation between the lowest accessible temperature $T_{\min}$ and the maximal holographic coordinate $z_{\max}$ depends on the specific form of $A(z)$. Thus, the mapping of thermodynamic quantities along the holographic direction is not fixed but varies with $A(z)$. Methods that regard $A(z)$ as fixed must reserve a large buffer in the holographic coordinate to accommodate all possible mappings, incurring unnecessary computational cost. In our approach, $z_{\max}$ is computed adaptively for the target $T_{\min}$ at each iteration~\cite{cai2024neural}, ensuring that the depth of the global network remains constant. The model therefore adjusts automatically to the lattice data without wasting computational resources.

\textbf{Optimization Objective.} Once the neural network architecture is fixed, the next step is to choose an appropriate loss function. In our setup, the total loss consists of two components,
\begin{equation}
    L = L_{\text{LQCD}} + L_{\text{R}},
\end{equation}
where $L_{\text{LQCD}}$ quantifies the deviation between the model predictions and the LQCD data, and $L_{\text{R}}$ represents regularization terms associated with physical constraints. These constraints include the Stefan–Boltzmann limit, the scalar-field equation of motion, and the positivity of $f$. We explicitly introduce these constraints into the designs of neural networks to enforce the regularization terms to vanish exactly. The only explicit model-regularization term is $L_{\text{R}} = L_{\text{AdS}} = N(A(0)-0)^2,$ which measures the degree to which the network satisfies the AdS boundary condition. $N$ is the total number of grid points along the holographic direction, and it keeps the two loss functions $L_{\text{LQCD}} $ and $ L_{\text{R}}$ at the same order of magnitude. A detailed explanation of this \textit{physics-driven deep learning} method~\cite{aarts2025physics} will be given in the next section. 

The first term $L_{\text{LQCD}}$ is defined as the mean square error(MSE) between the model predictions and the LQCD data for the EoS and baryon number susceptibility,
\begin{equation}
    L_{\text{LQCD}}= \left\{L_{s} , L_{\chi}\right\}.
\end{equation}
Here, entropy density error $L_{s}= \Sigma (s_{\text{NN}}-s_{\text{LQCD}})^2$, where NN denotes the values given by the neural network model, and LQCD denotes the LQCD data. $L_{\chi}= \Sigma (\chi^B_{2\text{NN}}-\chi^B_{2\text{LQCD}})^2$ represents the baryon number susceptibility error. Since there are only two independent quantities in the thermodynamic quantities, it is unnecessary to include additional errors. In addition, our optimization strategy trains $s$ and $\chi^B_2$ independently. This is because the equation of state at zero baryon number density is determined solely by the metric component $A$. In contrast, the baryon number susceptibility depends on both $A$ and the coupling function $f$. Therefore, a more reasonable approach is to first determine $A$ from the EoS (with the loss function $L_s + L_{\text{AdS}}$), and then extract $f$ using the baryon number susceptibility data(with the loss function $L_\chi$), rather than attempting to determine $A$ simultaneously from both datasets.
Furthermore, because there is no significant variation in the uncertainties within each dataset, we adopt the MSE loss.

Finally, the model is optimized using the Adam($lr=1e-3, betas=(0.9, 0.999), eps=1e-8$) optimizer, which is a popular choice for training neural networks. The optimization process adjusts the parameters of the neural networks representing $A$ and $f$ to minimize the loss function.

\section{constraints}
Our model is subject to certain physical and algorithmic constraints. Choosing an appropriate model architecture that naturally satisfies these constraints can improve accuracy and avoid the optimization complexities that arise when incorporating the constraints into the loss function. The main constraints incorporated in our model and their corresponding solutions are as follows. Among these, the $A(z)$ constraint and the Stefan–Boltzmann limit are physical constraints, while the constraints on the dilaton field and the coupling function are intrinsic to the model itself.

\textbf{$A(z)$ constraint.} To ensure that the model satisfies the AdS boundary conditions, we impose the following constraints:
\begin{equation}
A(0)=0.
\end{equation}
Meanwhile, empirically, the function $A(z)$ should be a monotonically decreasing function~\cite{zhang2022spectra}. Therefore, we use the \textit{softplus} activation function in the last layer, which helps the neural network architecture as much as possible conform to this prior behavior.

\textbf{Stefan-Boltzmann limit.} In some other discussions, the five-dimensional Newton's constant $G_5$ can also be treated as an adjustable parameter in the optimization process and, in principle, can be inferred from boundary LQCD data~\cite{Chen:2024ckb}. However, in our model, $G_5$ essentially serves the sole purpose of fixing the Stefan-Boltzmann limit at high temperatures. Admittedly, we cannot guarantee the validity of the holographic model in the extremely high-temperature regime, nor can we assert whether it should strictly adhere to the AdS/CFT duality limit. Nevertheless, since $s/T^3$ generally characterizes the number of effective degrees of freedom in the system, and the Stefan-Boltzmann limit from AdS/CFT, $(s/T^3)_{\text{AdS/CFT}}$, significantly overshoots the LQCD result $(s/T^3)_{\text{LQCD}}$ at high temperatures, this discrepancy arises because the AdS/CFT duality limit contains far more degrees of freedom than those captured by the 2+1 flavor LQCD data. Therefore, a reasonable choice is to fix $G_5$ such that the Stefan-Boltzmann limit matches that of LQCD.

In the UV limit as $z_H \to 0$, the temperature and entropy density can be easily obtained as
\begin{equation}
    \lim_{z_H \rightarrow0} T=\frac{1}{\pi z_H},\lim_{z_H \rightarrow0} s=\frac{1}{4 G_5 z_H^3}.
\end{equation}
The Stefan-Boltzmann limit of LQCD is given by~\cite{Borsanyi:2013bia}:
\begin{equation}
    \lim_{z_H \rightarrow0} \frac{s}{T^3}=\frac{\pi^3}{4 G_5}=4 \times  5.209.
\end{equation}
That is $G_5 = 0.372028$.

\textbf{Derivative behavior of dilaton field.} Motivated by prior knowledge, we employ the \textit{-softplus} activation function at the outermost layer of the $A$ network, which ensures that the following constraint is automatically satisfied in most cases,
\begin{equation}
-6\left(A^{\prime \prime}-A^{\prime 2}+\frac{2}{z} A^{\prime}\right)>0.
\end{equation}
The \textit{-softplus} function naturally enforces this inequality, effectively embedding the constraint into the network architecture. Consequently, no additional constraint term is required in the loss function, simplifying the overall loss design.

Moreover, neural networks typically prioritize optimizing low-frequency modes during training~\cite{xu2018training}. Since this model relies on higher-order derivatives, it is important to suppress the interference caused by high-frequency components at early stages and to ensure the smoothness of the functions, avoiding unphysical high-frequency degrees of freedom. To achieve this, the \textit{softplus} activation function can be employed to attenuate the high-frequency modes of the neural network effectively. Of course, this suppression does not compromise the expressive power of the neural network. If necessary, after sufficiently long training, the influence of the earlier hidden layers can still enable the network to capture localized high-frequency patterns.
\begin{figure*}[!htbp]
    \centering
    \includegraphics[width=0.45\linewidth]{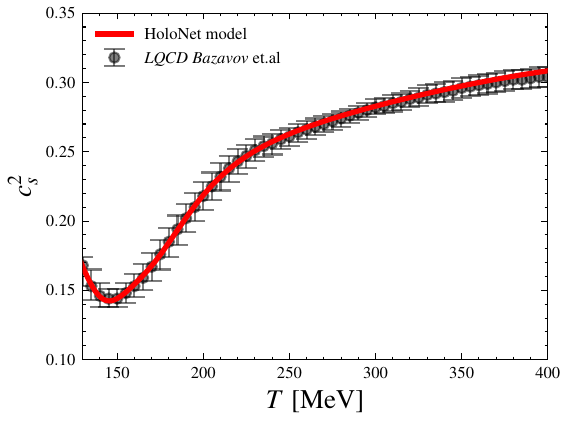}
    \includegraphics[width=0.45\linewidth]{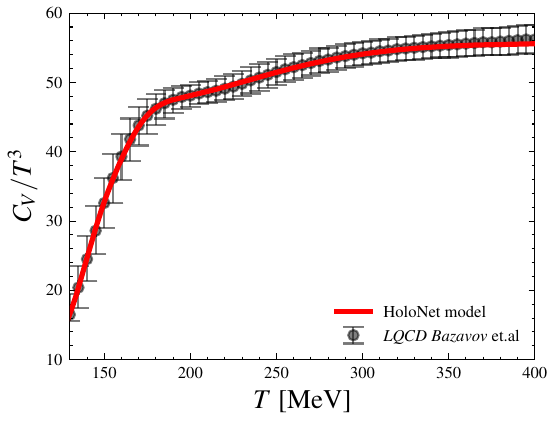}
    \caption{The figures display the speed of sound and specific heat, with red solid lines representing our HoloNet results and the error bars indicating the lattice data~\cite{bazavov2014equation}. As anticipated, these quantities exhibit excellent agreement once the model accurately satisfies the EoS.}
    \label{fig:cs2andCv}
\end{figure*}

\textbf{Positive-defined coupling function.} In general, we impose no strong restrictions on the coupling function $f(z)$, except that the following integral requires $ f(z) $ to be positive definite,
\begin{equation}
\int_0^{z_H} \frac{y}{e^{A(y)} f(y)} d y.
\end{equation}
This is because this term appears in the denominator in~(\ref{intform}) and therefore must not vanish (If it becomes zero, it would cause the baryon number susceptibility to diverge). A sufficient condition to ensure this is to require the function $f$ to be positive, which prevents cancellations between positive and negative contributions in the integral. The \textit{softplus} layer can effectively ensure this.

\section{Reconstruction Results}
The $(2+1)$-flavor LQCD data are taken from ~\cite{bazavov2014equation,borsanyi2012fluctuations}. From the LQCD data, through the optimization of the neural networks, we can reconstruct the bulk functions $A(z)$ and $f(z)$, as shown in Fig.~\ref{fig:Aandfform}.

\subsection{Thermodynamic Observables}
\begin{figure}[!hbtp]
    \centering
    \includegraphics[width=0.9\linewidth]{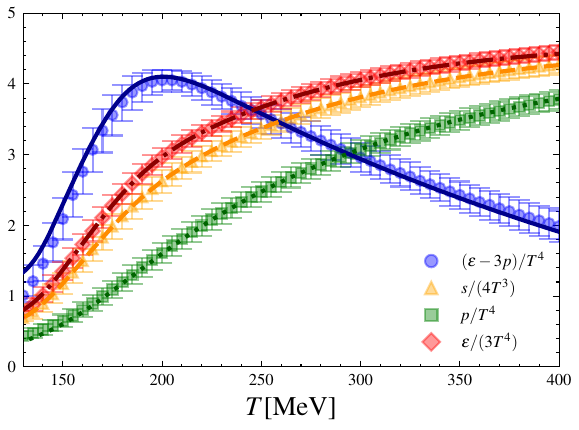}
    \caption{EoS comparison between neural network results and LQCD data~\cite{bazavov2014equation}. The error bars represent the LQCD data, while the solid lines represent the HoloNet model results. The blue curve corresponds to the trace anomaly $\epsilon-3p$, the yellow to the entropy density $s$, the green to the pressure $p$, and the red to the energy density $\epsilon$. The entropy density is used in training the HoloNet model. Since these are non-independent thermodynamic variables, the pressure, trace anomaly, and energy density generally agree with the lattice data automatically.}
    \label{fig:eos}
\end{figure}

Fig.~\ref{fig:eos} compares the LQCD data (error bars) at zero chemical potential with the optimized HoloNet results (solid lines). The comparison covers four quantities: energy density, entropy density, pressure, and anomaly. Among these, the entropy density is used to optimize the network, while the energy density, pressure, and anomaly automatically match the lattice results after optimization. At zero chemical potential, all these quantities depend only on the metric function $A(z)$.

\begin{figure}[!hbtp]
    \centering
    \includegraphics[width=0.9\linewidth]{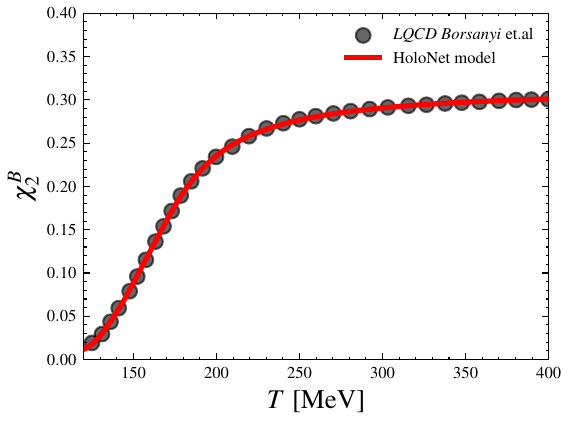}
    \caption{Baryon number susceptibility comparison between HoloNet result and LQCD data~\cite{borsanyi2012fluctuations}. The black dots represent the LQCD results, with error bars omitted due to their negligible size. The red solid line represents the neural network model results. The optimized model exhibits a high level of quantitative agreement with the lattice results.}
    \label{fig:chi-b}
\end{figure}

The baryon number susceptibility is shown in Fig.~\ref{fig:chi-b}. It is jointly determined by the metric function $A(z)$ and the coupling function $f(z)$. Here, $A(z)$ is treated as a completely external parameter with respect to the baryon number susceptibility. The susceptibility is used to optimize $f(z)$, allowing the model to be extended to finite chemical potential.

As the EoS is accurately satisfied, the speed of sound and specific heat are also well reproduced. The related results can be found in Fig.~\ref{fig:cs2andCv}.

\subsection{Closed-Form Expression}
In addition, we present an approximate closed-form expression. This expression is derived by first employing the \textbf{PySR(python)} package~\cite{cranmer2023interpretable} to perform symbolic regression and identify potential functional candidates, followed by manual inspection and refinement to propose a suitable analytical form for reference. That is,
\begin{equation}
\begin{aligned}
& A(z)=a_1 z^4-a_2 \log \left(1+z^2\right)-a_3 \log \left(1+a_4 z^4\right),
\\&  f(z)=f_1\  \text{sech}(f_2(z+f_3)^3).
\end{aligned}
\end{equation}
The corresponding constants are $a_1=0.00037885$, $a_2=0.062182$, $a_3=0.21002$, $a_4=0.020314$, $f_1=0.31197$, $f_2=0.079030$, $f_3=0.34070$.

\begin{figure}[!htbp]
    \centering
    \includegraphics[width=0.9\linewidth]{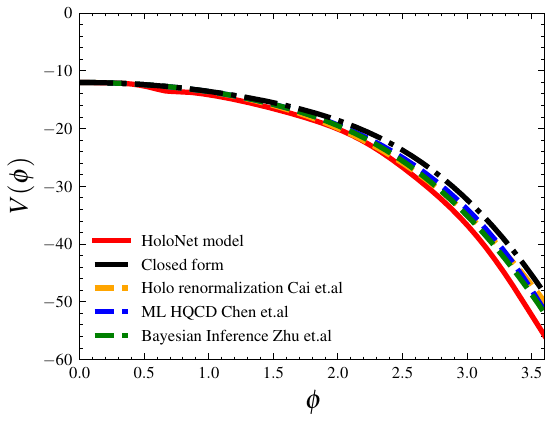}
    \caption{Reconstructed potential function $V(\phi)$ from HoloNet training(red). By contrast, the black line is an analytical closed-form result. The yellow dashed line represents the potential obtained from the holographic renormalization method, while the blue and green dashed lines represent the results from~\cite{chen2025flavor} and~\cite{zhu2025bayesian}, respectively. Within the $\phi$ range used in the numerical calculations, the values of $V(\phi)$ show a high degree of agreement.}
    \label{fig:v-phi}
\end{figure}

\begin{figure}[!htbp]
    \centering
    \includegraphics[width=0.9\linewidth]{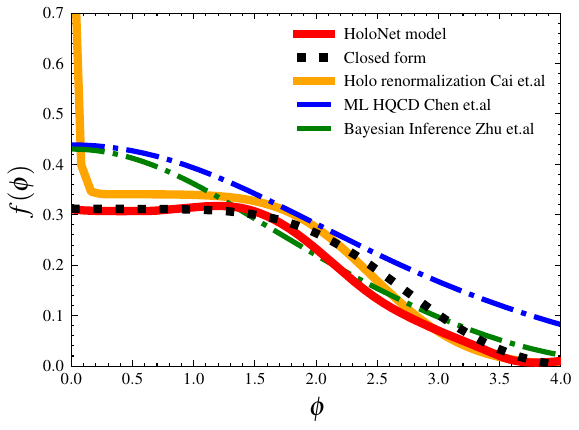}
    \caption{Coupling function $f(\phi)$ from neural network training(red). The black line is an analytical closed-form result. The yellow solid line represents the coupling function obtained from the holographic renormalization method, while the blue and green dashed lines represent the results from~\cite{chen2025flavor} and~\cite{zhu2025bayesian}. Within the $\phi$ range involved in the numerical calculations, the values of $f(\phi)$ are in close agreement with the holographic renormalization method.}
    \label{fig:f-phi}
\end{figure}

The analytic and neural-network results agree very well, with the analytic curves shown in black in Figs.~\ref{fig:Aandfform}, \ref{fig:v-phi}, \ref{fig:f-phi}, and \ref{fig:CEPplot}. In the temperature range 130–400 MeV, the maximal relative discrepancies between the neural-network reconstruction and the LQCD data are of order $\epsilon_s \sim 10^{-5}$ for the entropy density and $\epsilon_{\chi_2^B} \sim 10^{-7}$ for the baryon number susceptibility. For the analytic reconstruction, the corresponding discrepancies are slightly larger, $\epsilon_s \sim 10^{-4}$ and $\epsilon_{\chi_2^B} \sim 10^{-5}$.


We can reconstruct the potential functions $V(\phi)$ and $f(\phi)$ using the learned $A(z)$ and $f(z)$. The results are shown in Fig.~\ref{fig:v-phi} and Fig.~\ref{fig:f-phi}. The reconstructed potential functions exhibit quantitative agreement with those obtained through the holographic renormalization method~\cite{Cai:2022omk}. It should be noted that in the $\phi \to 0$ region, the holographic renormalization scheme introduces a pole in $f(\phi)$, leading to a slight difference between the two approaches. This pole, in ~\cite{Cai:2022omk}, was artificially introduced for computational convenience by making $f(0)=1$, has no physical significance, and can be removed. This consistency serves as a crucial validation of our neural network approach. Eventually, our reconstructions support the following analytical closed forms of $V(\phi)$) and $f(\phi)$,
\begin{equation}
\begin{aligned}
&V(\phi)=-12 \  \cosh(c_1 \phi)+c_2\phi^2,\\
&f(\phi)=c_3\  \text{sech}(c_4(\phi+c_5)^3),
\end{aligned}
\end{equation}
where $c_1=0.69048$, $c_2=1.2777$, $c_3=0.31562$, $c_4=0.061371$, $c_5=0.35482$.

\subsection{Extrapolation to CEP}

\begin{figure}[!htb]
    \centering
    \includegraphics[width=0.9\linewidth]{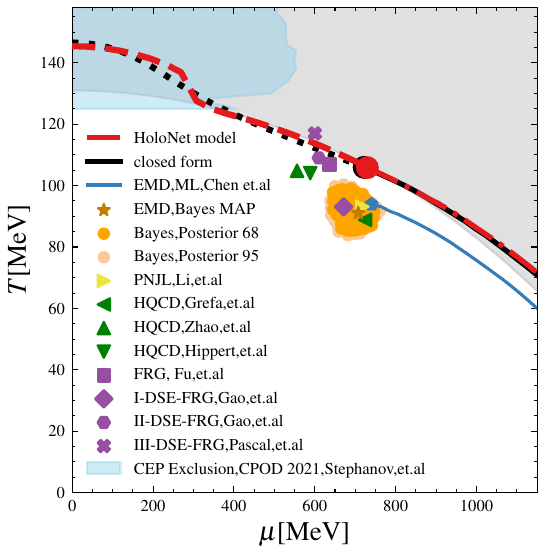}
    \caption{The phase diagram is shown, where the red solid and dashed lines denote the first-order phase transition and crossover obtained from the neural network, and the black line indicates the analytical guess. The gray-shaded region represents the range in which the model predictions are reliable, obtained by extending the zero-chemical-potential equation of state to finite chemical potential through the baryon number susceptibility in the holographic model. The light-blue region indicates the part of the phase diagram where the CEP has been excluded \cite{borsanyi2025lattice}. Green ones represent DeWolfe-Gubser-Rosen-type HQCD models, and green left triangle \cite{grefa2021hot}, upward triangle \cite{zhao2023phase}, and downward triangle \cite{hippert2024bayesian} correspond to related HQCD results. Blue dot and solid line denote the CEP and first-order lines from \cite{chen2025flavor}. The orange and light orange regions show the $68\%$ and $95\%$ confidence-level CEP estimates from \cite{zhu2025bayesian}, with the dark orange pentagram indicating the maximum a posteriori result. The yellow right triangle is the result of PNJL \cite{li2019kurtosis}. Purple square, and purple diamond represent the CEPs obtained from the FRG \cite{fu2020qcd}, and DSE-FRG \cite{gao2020qcd} approaches, respectively. The results of the further DSE–FRG approach are shown as orange hexagons~\cite{gao2021chiral} and purple crosses~\cite{gunkel2021locating}.}
    \label{fig:CEPplot}
\end{figure}

We have also computed the locations of the CEP, the first-order phase transition line, and the crossover in the phase diagram in Fig.~\ref{fig:CEPplot}. Overall, our results are somewhat higher in temperature compared with most previous studies. Fortunately, these results fall within the region represented by our data, making them numerically reliable to a certain extent. Nevertheless, as they lie close to the boundary, we strongly recommend treating them merely as a reference. Access to lattice data at lower temperatures would enable further extrapolation and yield more robust and trustworthy results.

Finally, it is necessary to clarify an important point. The LQCD data employed here are restricted to a specific temperature range, which naturally corresponds to a certain region of $z$. This region is indicated by the gray-shaded area in Fig.~\ref{fig:Aandfform}, and our results are considered valid only within this range (derived from the neural network results). Upon extension to finite chemical potential, there exists a corresponding region in the $T-\mu$ plane, highlighted in gray in Fig.~\ref{fig:CEPplot}. Other unmarked regions are either irrelevant to this consideration or entirely fall within the valid range.

\section{Conclusion and Discussion}
In this work, we construct a fully data-driven potential-reconstruction holographic QCD model, HoloNet, where the metric components and coupling function are represented by two independent neural networks embedded along the holographic direction, eliminating any need for assumed functional forms. To maintain self-consistency without increasing model complexity, mild assumptions are imposed only on the final layers so that certain constraints are approximately satisfied, without restricting the model’s expressive power. The network is trained on 2+1 flavor LQCD data for the entropy density and baryon number susceptibility, achieving very small errors, while other thermodynamic quantities—energy density, pressure, trace anomaly, speed of sound, and specific heat—naturally match the lattice results through thermodynamic relations. Since the lattice data span 130–400 MeV, the model provides reliable predictions within this temperature range.

We have also computed the QCD phase diagram and identified a CEP at $(T = 106\ \text{MeV},\ \mu = 730\ \text{MeV})$ in our model. Although the functions $A(z)$ and $V(\phi)$ are similar to those used in many existing approaches, the CEP position is predominantly controlled by $f(\phi)$ and is highly sensitive to its precise form. As a result, our predicted CEP exhibits a noticeable deviation from most earlier results. Additional uncertainties arise because the equations of motion involve higher-order derivatives of $A(z)$ and $f(z)$, whose accuracy decreases near the edge of the region where the data can be reliably extrapolated. This limitation introduces further uncertainty in the CEP determination, though the results in these regions still provide a meaningful reference. Extending lattice data to slightly lower temperatures would enlarge the model’s reliable domain and is likely to bring the CEP prediction into closer alignment with values commonly reported in the literature, offering a promising direction for future work.

Another issue worth discussing is temperature dependence. Since $g(z)$ depends on $z_H$, the values of $V(z)$ in the holographic coordinate also depend on $z_H$, which suggests that the form of $V(z)$ may be related to the specific temperature value. Meanwhile, $\phi$ appears to depend only on our $A$ neural network and is independent of temperature, so $V$ seems to be not only a function of $\phi$, $V(\phi)$, but may also explicitly depend on temperature, $V(\phi, T)$. However, in practice, after incorporating the lattice data, the temperature dependence of $V$ disappears. Numerically, it remains consistent with the form of $V(\phi)$ in the action, i.e., $V$ depends only on $\phi$. Relevant discussions can be found in the Appendix.

Finally, we compared our results with those obtained via the holographic renormalization method~\cite{Cai:2022omk} and the potential-reconstruction approach~\cite{chen2025flavor} based on a prescribed ansatz for $A(z)$. These traditional methods generally differ because holographic renormalization assumes a specific form for $V(\phi)$, while potential reconstruction assumes a parameterized $A(z)$. In contrast, our fully machine-learning reconstruction makes no such ansatz yet reproduces results consistent with holographic renormalization, indicating that the two approaches are in fact equivalent. We also find that the pole of $f(\phi)$ at $z=0$ that appears in holographic renormalization has no significant effect on the equation of state or the CEP location and can be regarded as a gauge choice. Since both potential reconstruction and holographic renormalization derive from the same action, their agreement provides an important validation of the self-consistency of the EMD framework. Geometrically, as noted earlier, the temperature $T$ may be encoded in either the warp factor $A(z)$ or the horizon position $z_H$; the latter is fixed in holographic renormalization for numerical convenience, whereas our model allows it to vary. Because fixed and unfixed $z_H$ spacetimes are related by diffeomorphisms, the consistency between our reconstructed potential and that from holographic renormalization further confirms the internal coherence of the framework.

\section*{Acknowledgements}
We thank Drs. Daichi Takeda, Zhibin Li, and Jan M. Pawlowski for helpful discussions.
We thank the DEEP-IN working group at RIKEN-iTHEMS for support in the preparation of this paper.
LW is supported by the RIKEN-TRIP initiative (RIKEN Quantum), JSPS KAKENHI Grant No. 25H01560, and JST-BOOST Grant No.JPMJBY24H9; MH is supported in part by the National Natural Science Foundation of China (NSFC) Grant Nos: 12235016, 12221005. HZ is supported by the Science and Technology Development Plan Project of Jilin Province, China Grant No. 20240101326JC.


\bibliographystyle{unsrt} 
\bibliography{ref} 

\end{document}


\onecolumngrid

\appendix 
\clearpage
\renewcommand{\appendixname}{}

\section*{Supplementary material}

\section{Appendix A: temperature-independent potential}

Our starting point is a temperature-independent potential $V(\phi)$. However, the result we obtain is an explicitly temperature-dependent potential $V(\phi, T)$. This discrepancy arises from the assumption that all black hole solutions share the same warp factor $A$, under which this temperature dependence naturally emerges from the equations of motion. Once this assumption is imposed, $A$ no longer carries any temperature dependence, while $V(z)$ does, making it impossible to cancel the temperature dependence that appears in $V(\phi)$.
To strictly remove this temperature dependence, one cannot use the same $A(z)$ function for every solution. Instead, the warp factor must be solved independently for each background. This is precisely the strategy employed in works such as \cite{rougemont2024hot, Cai:2022omk}.

\begin{figure}[!h]
    \centering
    \includegraphics[width=0.6\linewidth]{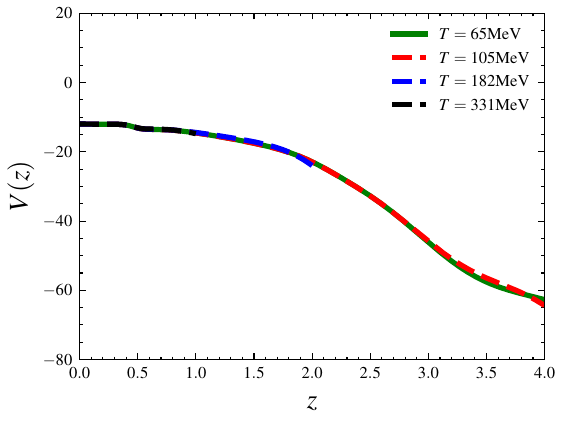}
    \caption{The green, red, blue, and black curves in the figure correspond to temperatures of 65 MeV, 105 MeV, 182 MeV, and 331 MeV, respectively. In the region where lattice data are available (above about 130 MeV), these curves are almost indistinguishable. This indicates that the potential $V(z)$ consistent with the lattice data exhibits no temperature dependence, and consequently, the potential $V(\phi)$ is also temperature independent.}
    \label{fig:differentT}
\end{figure}

This phenomenon is quite intriguing. What happens if we feed the lattice data into such a model? The result, as shown in Fig.~\ref{fig:differentT}, is remarkable: the model that fits the lattice data automatically eliminates the temperature dependence. In the figure, the influence of temperature on $V(z)$ is minimal, and the noticeable deviation occurs only below 130 MeV—outside the range covered by the lattice data, and thus not discussed further.

This implies that, in fact, we initially started from a temperature-dependent model, but through the optimization process, it was automatically iterated into a temperature-independent one. Therefore, the result is self-consistent with the potential as defined in the action. This is highly nontrivial, as the model imposes no soft or hard constraints in this regard; the behavior emerges solely from the information in the lattice data.

\bibliographystyle{unsrt} 
\bibliography{ref} 